\documentclass[a4paper,11pt]{article}
\pdfoutput=1 % if your are submitting a pdflatex (i.e. if you have
\usepackage{jinstpub} % for details on the use of the package, please
\usepackage{lineno} % line
\usepackage{multicol}
\usepackage{subcaption}
\usepackage{color}
\usepackage{ulem}
\usepackage{hyperref}
\usepackage{wrapfig}
\title{\boldmath Design and Development of the Core Software for STCF Offline Data Processing}

\author[a,b]{W. H. Huang }
\author[c,d]{H. Li }
\author[c,d]{H. Zhou}
\author[a,b]{T. Li }
\author[e]{Q. Y. Li }
\author[a,b,1]{X. T. Huang \note{Corresponding author.}}

\affiliation[a]{Shandong University,\\Qingdao,Shandong, 266237, China}
\affiliation[b]{Key Laboratory of Particle Physics and Particle Irradiation (MOE), Shandong University,\\Qingdao, Shandong, 266237, China}
\affiliation[c]{University of Science and Technology of China,\\HeFei, Anhui, 230026, China}
\affiliation[d]{State Key Laboratory of Particle Detection and Electronics
Department of Modern Physics,\\HeFei, Anhui, 230026, China}
\affiliation[e]{Shandong Institute of Advanced Technology,\\Jinan, Shandong, 250100, China}

\emailAdd{huangxt@sdu.edu.cn}

\abstract{

The Super Tau Charm Facility (STCF) is a proposed electron-positron collider working at $\sqrt{s}=2\sim 7$ GeV, and the peak luminosity is designed to be above $0.5 \times 10^{35}cm^{-2}s^{-1}$. The huge amount of scientific data brings great challenges to the offline data processing software, including the Monte Carlo simulation, calibration, reconstruction as well as the data analysis. To facilitate efficient offline data processing, the offline software of Super Tau Charm Facility (OSCAR) is developed based on SNiPER, a lightweight framework designed for HEP experiments, as well as a few state-of-art software in the HEP community, such as podio and DD4hep.\par
This paper describes the design and implementation of  the core software of the OSCAR, which provides the foundation for the development of complex algorithms to be applied on the large data sets produced by STCF, particularly the way to integrate the common HEP software toolkits, such as Geant4, DD4hep and podio, into SNiPER. The software framework also provides a potential solution for other lightweight HEP experiments as well.

}

\keywords{Core software; Offline data processing; STCF; OSCAR; SNiPER}

\collaboration[c]{}

\begin{document}
%\linenumbers  % line
\maketitle
\flushbottom

\section{Introduction\label{sec:intro}}
\label{sec:intro}

The Super Tau Charm Facility (STCF)~\cite{Peng:2020orp} is  a new-generation facility of electron positron collider operating at center-of-mass energies of  $2-7$ GeV, the transition region between perturbative and non-perturbative quantum chromodynamics. STCF will play a leading role in the tau, charm and hadron physics of high energy physics intensity frontier in the world. The peak luminosity of STCF is designed to be $0.5 \times 10^{35}cm^{-2}s^{-1}$, almost two orders of magnitude higher than the present Tau-Charm factory. Meanwhile, STCF is expected to produce huge amount of scientific data, which poses many challenges to the STCF detector, online data acquisition~(DAQ) as well as offline data processing. STCF is currently under extensive exploration with a technology research and development program covering design and optimization of  the STCF accelerator, detector, trigger, DAQ and offline software. To facilitate current detector conceptual design, physics potential studies as well as future raw data analysis, an offline software of Super Tau Charm Facility (OSCAR) is designed and developed to provide a common and high performance platform for STCF offline applications, including physics generators, detailed detector simulation, data calibration, event reconstruction as well as physics analysis.

\clearpage

\begin{wrapfigure}[15]{r}{0.5\linewidth}
\centering
\includegraphics[height=180pt, width=140pt]{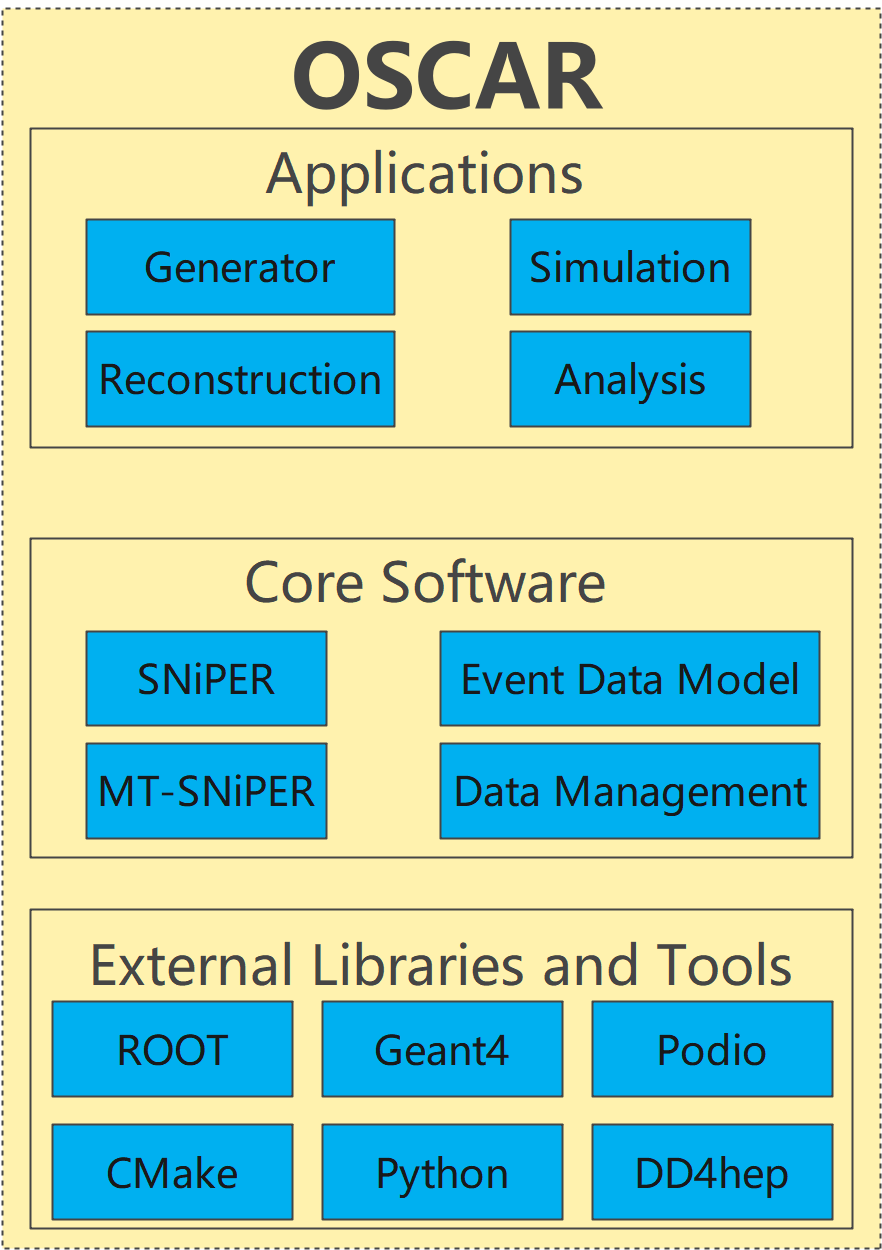}
\caption{\footnotesize The architecture of OSCAR.}
\label{fig:design}
\end{wrapfigure}

The architecture of OSCAR is depicted in Figure~\ref{fig:design}. It mainly consists of three layers: the bottom layer is for external libraries, the middle layer is for the core software and the top layer is for STCF specific applications. External libraries used by OSCAR contain the software and toolkits not only frequently used in HEP experiments, like ROOT~\cite{ref1} and Geant4~\cite{GEANT4:2002zbu}, but also some of the Common Turnkey Software Stack (Key4hep)~\cite{Ganis:2021vgv}, such as DD4hep~\cite{Frank:2014zya} and podio~\cite{Gaede:2021izq}, which are adopted for the detector description and the event data model, respectively. The core software mainly consists of an underlying framework, parallel computing supports, event data model and a data management system. It provides key and common functionalities for the whole offline data processing. Experiment specific applications refer to the software of physics generators, detector simulation, digitization, calibration, reconstruction as well as physics analysis, which are developed based on the functionalities and interfaces provided by the core software. Therefore, the core software plays very crucial roles for performance of OSCAR.

{This paper aims to present the design and implementation of the core software of OSCAR. Section~\ref{sec:intro} gives a brief introduction to the offline software  system. Section~\ref{sec:req} outlines the challenges and requirements of STCF offline software. In Section~\ref{sec:imp},  the implementation of the core software of OSCAR is described in more detail, including the SNiPER framework and its parallel computing supports, the event data model, the management of event data as well as detector data. Section~\ref{sec:perf} presents some performance of applications developed based on the core software. Finally, Section~\ref{sec:conc} summarizes the current status and gives an outlook on the development of OSCAR.}

\section{Requirements and Design\label{sec:req}}

\subsection{Requirements and Challenges}

\indent{The core software is the base of the whole offline software. It builds the skeleton of the offline software for all application developers and physicists to develop their application algorithms, services and tools for event generators, detector simulation,  event reconstruction and physics analysis. Furthermore, the wide center-of-mass energy covering region, high peaking luminosity as well as high readout channel density of the STCF accelerator and detector pose severe challenges for the core software of OSCAR.}

$\bullet${High event rate and huge data volume. The center-of-mass energy region of STCF spans from 2 to 7 GeV and event rates vary with different center-of-mass energies. The highest one is estimated to be approximately 400 kHz at $J/\psi$ peak. In one operating year, STCF will collect the order of $10^{12}$ $J/\psi$ data sample. In addition, the average event size is expected about 45.7 KB. Therefore, the total data volume of raw data could reach the order of 100 PB per year. All of the data produced in one year should be fully reconstructed in less than six months for physics analysis.  Furthermore, at least the same size of the Monte Carlo~(MC) data will be produced for systematical uncertainty study.  So the core software should adopt versatile technologies to meet these requirements and challenges.\par}
$\bullet${Long timescale and high flexibility. The STCF team has completed the preliminary conceptual design and delivered the Conceptual Design Report at the end of 2022. Now the key technology research and development projects are expected to be finished and a Technology Design Report will be released by the end of 2025. STCF is expected to start commissioning in 2031-2033 after 5-7 years' construction and to keep operating for 10-15 years. During such a long timescale, OSCAR serves as the key software system for the offline data processing and physics analysis. So the core software should have good friendliness for developers and physicists as well as high flexibility for new technologies, including machine learning and parallel computing.\par}

\subsection{Overall Design}

\indent{To meet the requirements and challenges above, the core software of  the OSCAR is designed with following key criteria:}

$\bullet${ Parallelism. Given the huge data volume produced at STCF each year, parallelism is one of the necessary options for the OSCAR to speed up the data processing and meet the physics requirements. As a kernel part of the OSCAR, the core software provides key functionalities of event data processing mechanism, data management as well as Input/Output~(I/O) system. Therefore, parallel computing should be well supported with versatile technologies.\par}

$\bullet${ Modularity and Low Coupling. OSCAR consists of lots of components and packages for different applications which evolve with the design, construction and operation of STCF. The core software should provide mechanisms to ensure modulariy of OSCAR and low coupling between different components. In this way, every component takes care of specific functionality and provides well defined interfaces by which different components interact with each other. At the same time, key components should be independent with each other and shield the implementation of certain functionality from technologies adopted by OSCAR.\par}

$\bullet${ Flexibility, Stability and Compatibility. Given the long lifetime of STCF, the core software should provide standard interfaces for developers and physicists to easily plug in their applications in term of specific algorithms. Meanwhile, the impact of changes from new software or technologies should be minimized for users.\par}

\section{Key Components and Implementations\label{sec:imp}}

\begin{figure}[h]
\centering
\includegraphics[width=0.98\linewidth]{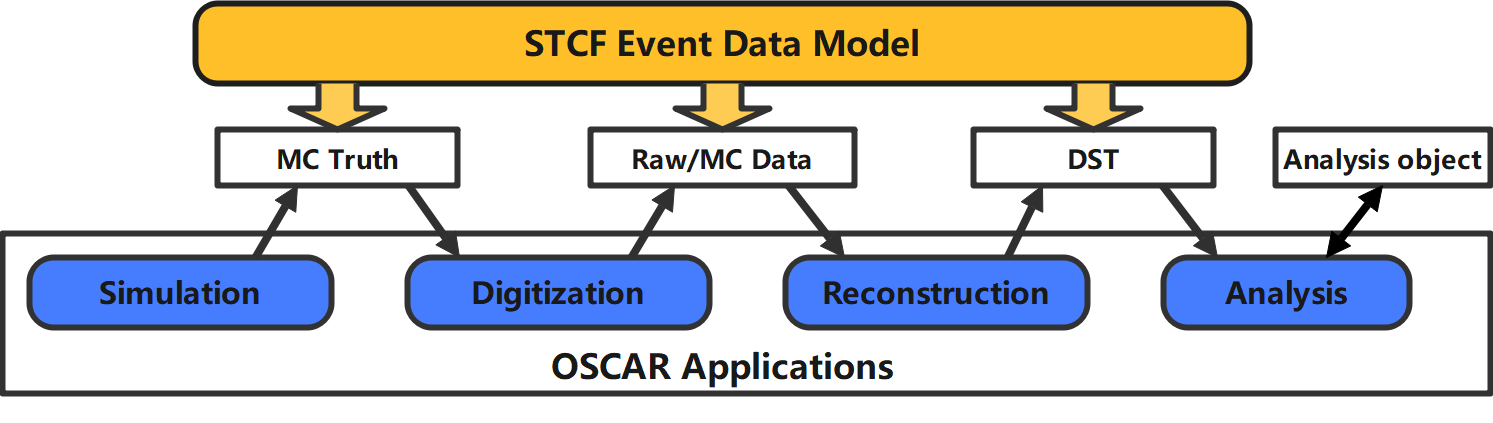}
\caption{\label{label}The STCF data processing applications and data flow.}
\label{fig:dataflow}
\end{figure}

\indent{The core software for STCF offline data processing adopts SNiPER as the underlying framework. Intel Threading Building Blocks (TBB)~\cite{ref2} based MT-SNiPER~\cite{Zou:2019cyq} provides well support for parallel computing within SNiPER. The data management system handles not only event data but also detector data. Together with OSCAR based applications, the event data model and event data management system form the key components of offline event data processing and data flow as shown in Figure~\ref{fig:dataflow}. The event data model defines event data objects for different applications including simulation, digitization, reconstruction and analysis, which are developed based on the interfaces of SNiPER framework. Every application is a producer and/or consumer of event data for other stages. All applications retrieve event data from the event data management system and register updated event data or new event data back to the event data management system. The event data in different application stages contains different kinds of event information as illustrated in  Figure~\ref{fig:datamodel}. It could be written out to the disk as the format of root files and be read from root files back in any stage. All the data files will be well managed by a file catalogue management system and would be accessed with the meta data by users. These key components and their implementations are described in more detail as following.}

\subsection{SNiPER Framework}

\indent{SNiPER is a lightweight software developed for HEP experiments and has been successfully used in a few experiments, such as Jiangmen Underground Neutrino Observation~(JUNO)~\cite{JUNO:2015zny}, Large High Altitude Air Shower Observation~(LHAASO)~\cite{LHAASO:2019qtb}, Neutrinoless double beta decay experiment~(nEXO)~\cite{nEXO:2018ylp} and so on. Distinguishing features of SNiPER are described as follows:}

\begin{wrapfigure}[10]{r}{0.5\linewidth}
\centering
\includegraphics[height=8pc, width=18pc]{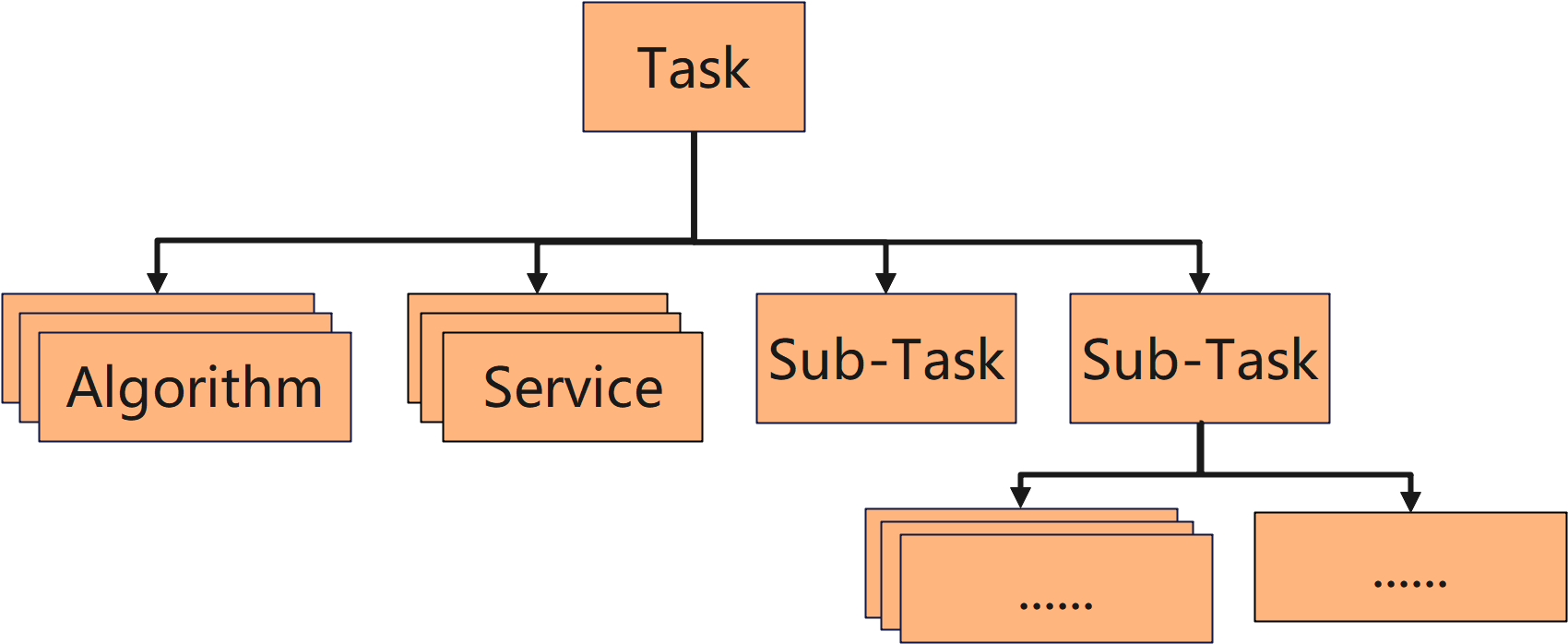}
\caption{\footnotesize Components trees in \emph{Task}s.}
\label{fig:task}
\end{wrapfigure}

\begin{figure}[h]
\centering
\includegraphics[width=0.75\linewidth, height=13pc]{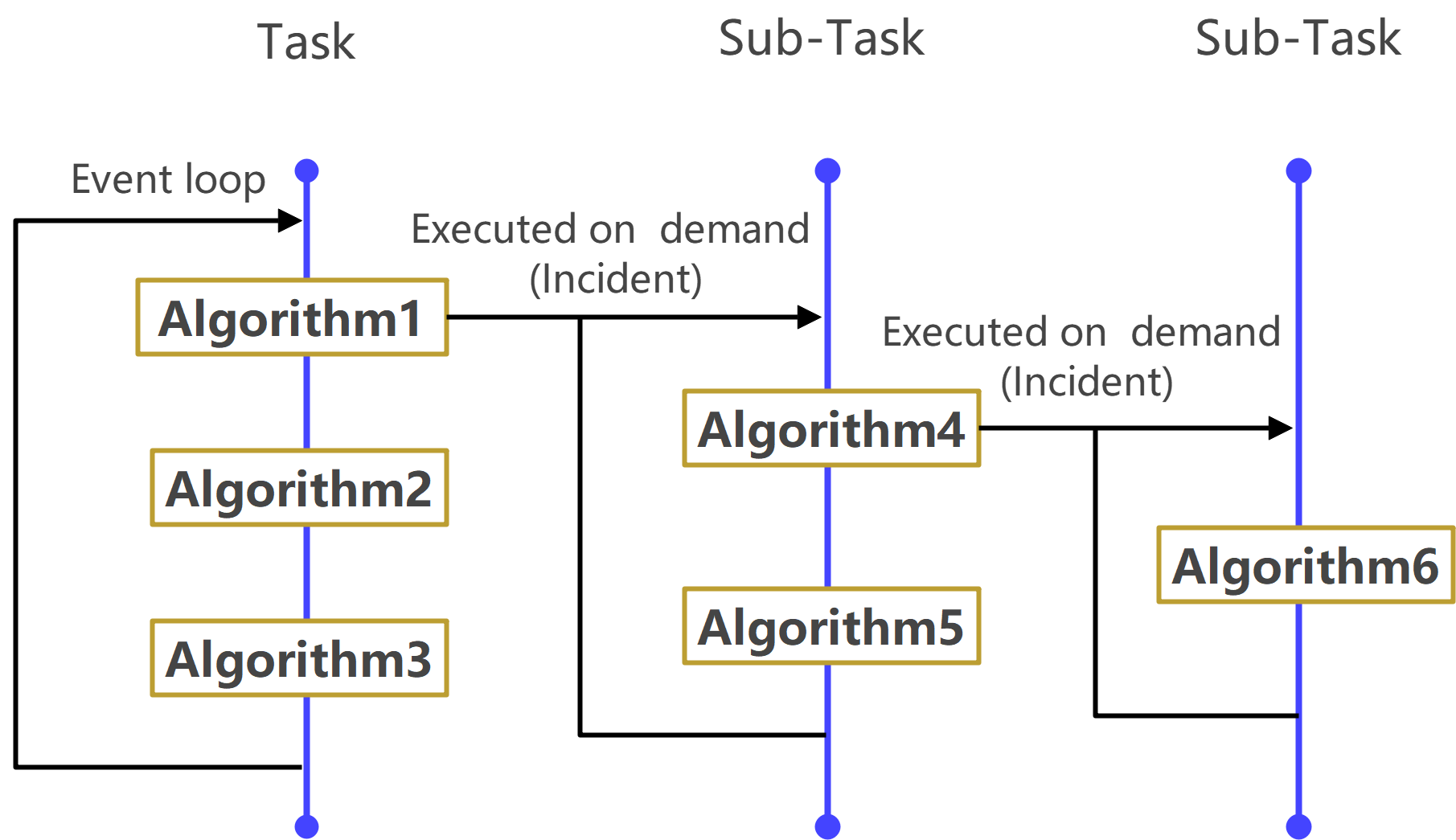}
\caption{\label{label}Conditional execution of multiple \emph{Task}s.}
\label{fig:sequence}
\end{figure}

$\bullet$ Support flexible event processing sequences. As shown in Figure~\ref{fig:task}, the unit of SNiPER is \emph{Task}, where developers can embed their \emph{Algorithm}s and \emph{Service}s to perform certain manipulation on event data. Each \emph{Task} instance has its own \emph{DataStore}, which handles even data in the memory and manages its lifetime. With this mechanism, every \emph{Task} controls its own event loop independently. Furthermore, SNiPER provides well support for multiple \emph{Task}s.  As shown in Figure~\ref{fig:sequence}, typical multiple \emph{Task}s are built with a nested \emph{Task} structure which can be easily configured in a single SNiPER job. \emph{Algorithm}s in one \emph{Task} would be executed in order and \emph{Algorithm}s in \emph{Sub-Task} should be executed on demand with the \emph{Incident} mechanism. For example, in an event mixing procedure, signal and background samples are produced independently. Then each sample can be configured as an input for each \emph{Sub-Task} instance. In this way, the event mixing algorithm in upper \emph{Task} triggers signal and background inputs on demand according to signal-to-noise ratio.

$\bullet$ {Support custom data management and I/O. SNiPER defines the abstract interfaces for applications to access to event data from the data management system and provides a novel mechanism to manage multiple events within certain time windows (called event buffering). In addition, SNiPER does not constrain implementations of data management and I/O system based on experiment specific requirements. For instance, a first input first output (FIFO) data buffer~\cite{Zou:2015ioy} is developed, which enables high efficiency data access and storage, as well as the capability to retrieve the events within a defined time window. FIFO data buffer stores adjacent events and provides a sophisticated method of memory updating. When turning off the time window, FIFO serves as a buffer which holds event data objects only for one independent event.}

$\bullet${Support parallel computing. With multiple \emph{Task}s and FIFO , SNiPER has intrinsic superiorities in parallel computing. To simplify the development of paralleled applications, MT-SNiPER is developed as a non-invasive wrapper of SNiPER kernel modules, so that it is almost transparent for developers and physicists to execute their applications in parallel without much modifications. The implementation of parallelism of the OSCAR will be further discussed in Section \ref{sec:para}.}

\subsection{Event Data Model and Data Management}

\indent{Event data model, serving as the heart part of offline software, does not only define event information and correlations between event data objects in different processing stages, but also provides interfaces and communication channels between different application algorithms as well. The data management system is the fundamental component for event data transmissions, communications between OSCAR based applications and data version control in offline data processing. Furthermore, the data management system should provide well supports for parallel computing throughout all stages of data. Therefore, event data model and data management are very essential and the most complicated components for the design and development of the STCF offline software system.}

\subsubsection{{Event Data Model Based on Podio}\label{sec:podio}}
\
\newline
\indent{During the offline data processing,  application algorithms would not directly use event data objects in the persistency store (persistent data), but instead, use event data objects in memory (transient data). In OSCAR, both persistent data and transient data are defined with one set of description for event data objects. Therefore, it is crucial to design an efficient event data model that meets all use cases in offline data processing and physics analysis.}\par 

\begin{figure}[h]
\center
\includegraphics[width=0.98\linewidth, height=20pc]{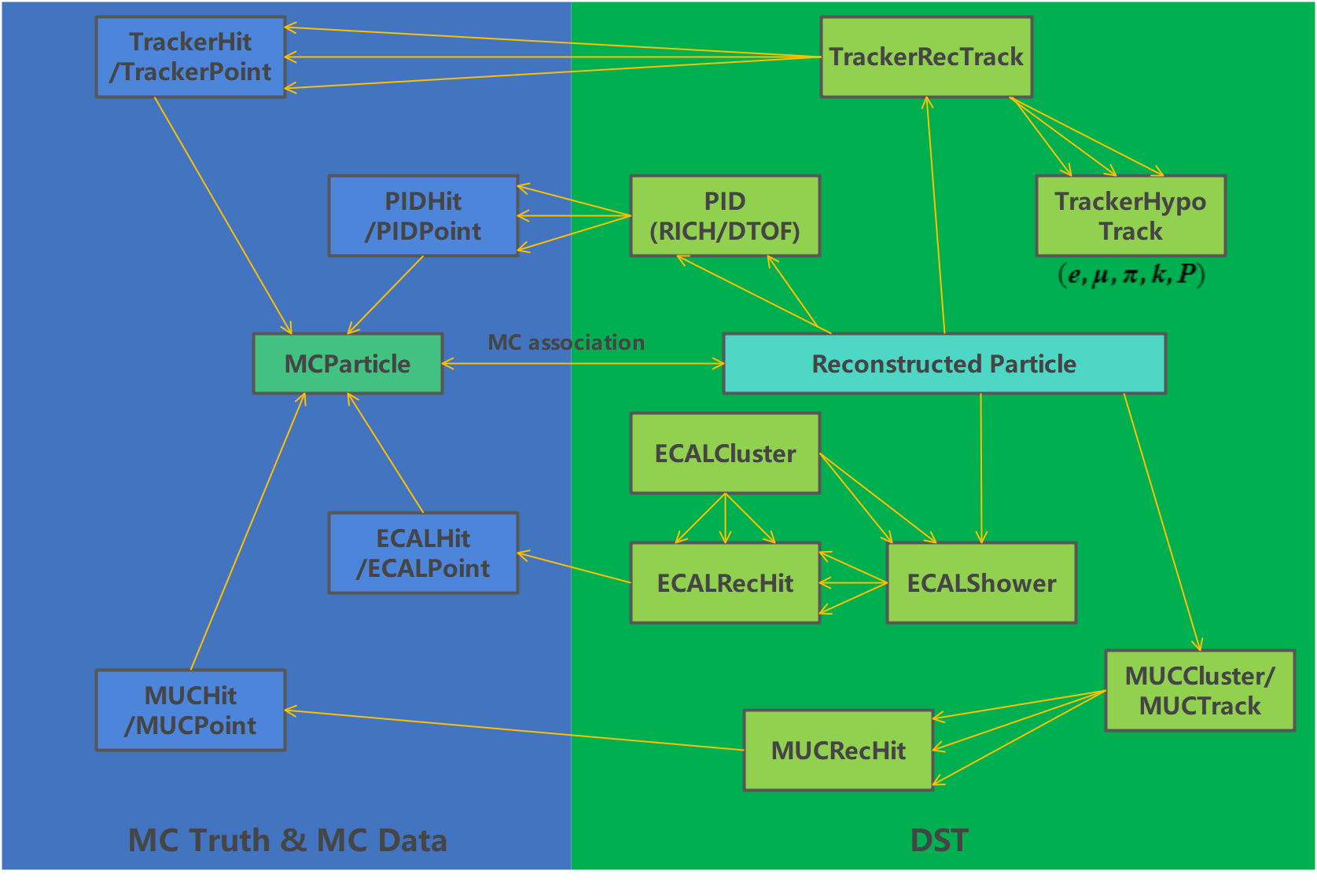}
\caption{\label{label}Podio based event data model for STCF.}
\label{fig:datamodel}
\end{figure}

{Traditional HEP event data models are heavily object oriented. These kinds of event data objects do have good encapsulation on event information and play very important roles in past decades. However, they are not so suitable for parallel computing. Now, a new C++ toolkit to define event data models with plain-old-data data structures and high efficient I/O performance~(podio) is emerging. Podio favors composition over inheritance and has well supports for parallel computing. The EDM4hep, aiming to design a common event data model for future collider experiments, has been developed based on podio and been adopted by Future Circular Collider~(FCC)~\cite{FCC:2018byv}, The Circular Electron Positron Collider~(CEPC)~\cite{CEPCStudyGroup:2018ghi} and some other experiments. The event data model for STCF is also developed based on podio. The event information and relationships between different event data objects are described in yaml~\cite{ref3} files, based on which C++ code for event data classes are automatically generated. According to compositions of the STCF detector, the classes for MC simulation~(MC Truth and MC Data) and reconstruction~(DST) are defined as described in Figure~\ref{fig:datamodel}. For instance, \emph{ECALShower}, \emph{ECALCluster}, \emph{ECALRecHit} are created for Electromagnetic Calorimeter~(ECAL) reconstruction and \emph{ECALHit}, \emph{ECALPoint} are created for ECAL simulation. For other sub detectors, like Inner Tracker, Main Drift Chamber~(MDC) and Particle Identification~(PID), corresponding classes are also created. Meanwhile, the relationships between different event data objects are implemented with ``one to many" or ``one to one" mechanism. In Figure~\ref{fig:datamodel}, single arrow between classes means they have the ``one to one" relationship while multiple arrows between classes represents the ``one to many" relationship. With this strategy and implementation, the event data model for STCF is naturally thread-safe and provides well supports for parallel computing.\par}

\subsubsection{Event Data Management for Serial Computing\label{sec:seria}}
\
\newline
\indent{Both SNiPER and podio have their own event data management system. Therefore, integrating the event data management of podio into OSCAR is a more vital and complicated procedure in order to use podio based event data model efficiently. The first step of the integration is adopting podio based event data model in the serial computing of OSCAR.}\par

\begin{wrapfigure}[16]{r}{0.5\linewidth}
\centering
\includegraphics[height=180pt, width=200pt]{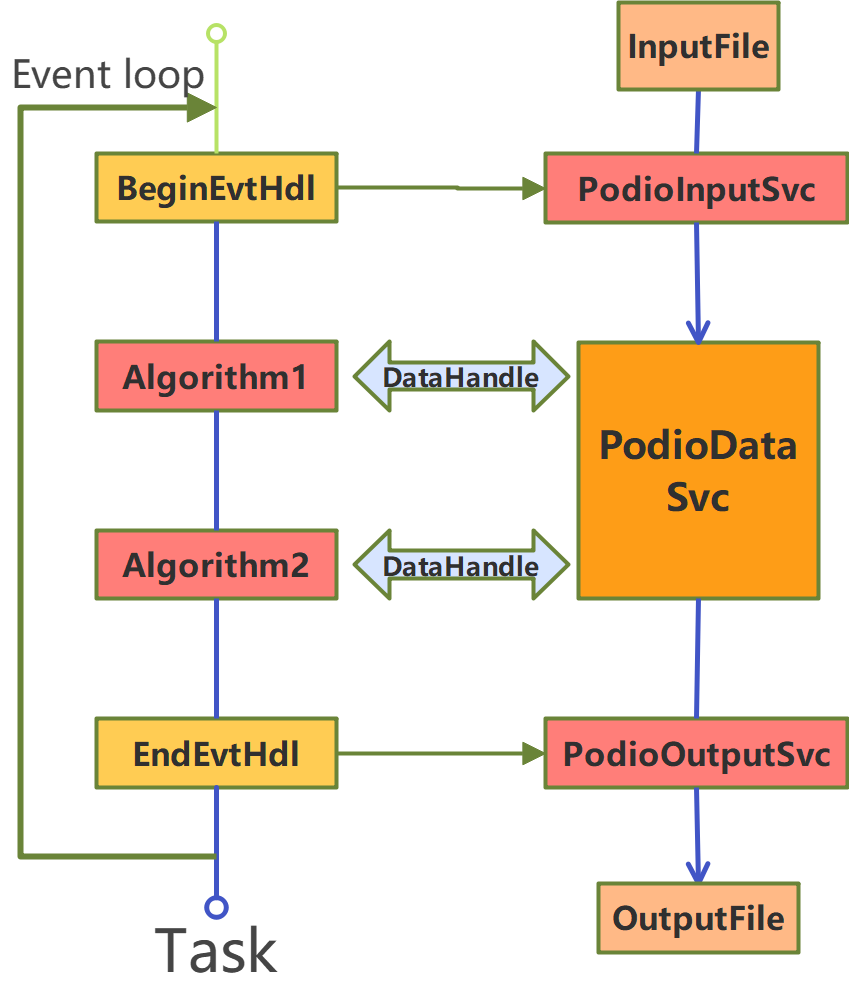}
\caption{\footnotesize Event data management for serial computing.}
\label{fig:datastoreinpodio}
\end{wrapfigure}

The event data management of podio mainly consists of three parts: \emph{EventStore} handles transient data, while \emph{ROOTReader} and \emph{ROOTWriter} manage data input and output, respectively. As illustrated in Figure~\ref{fig:datastoreinpodio}, \emph{PodioDataSvc}, \emph{PodioInputSvc}, \emph{PodioOutputSvc} and \emph{DataHandle} are developed to integrate podio within OSCAR. Based on \emph{EventStore} of podio, \emph{PodioDataSvc} is developed to manage transient event data within OSCAR. In one \emph{Task}, there is a pair of \emph{Incident}s named \emph{BeginEvtHdl} and \emph{EndEvtHdl}. Both \emph{BeginEvtHdl} and \emph{EndEvtHdl} are triggered automatically at the beginning and end of an event loop, respectively. Via \emph{PodioDataSvc}, \emph{PodioInputSvc} is invoked by \emph{BeginEvtHdl} to convert persistent event data from files into transient event data in \emph{EventStore}, while \emph{PodioOutputSvc} is invoked by \emph{EndEvtHdl} to perform conversion from transient event data in \emph{EventStore} into persistent event data in files. \emph{DataHandle} is developed as an user end template class to provide interfaces of identifying and accessing event data objects from \emph{PodioDataSvc}. In this way, the event data management system and event data model could match together and work in serial computing within OSCAR. Many simulation and reconstruction algorithms have been developed based on them.

\subsubsection{Event Data Management for Parallel Computing\label{sec:para}}
\
\newline
\indent{To speed up data processing and make full use of computing resources, parallel computing should be well supported by not only the underlying SNiPER but also the podio compatible OSCAR. MT-SNiPER is developed to support the parallelism of SNiPER based on Intel TBB. Multiple SNiPER Task Scheduler (\emph{Muster})~\cite{Zou:2019cyq} is one of key components of MT-SNiPER. When using \emph{Muster}, multiple instances of \emph{Task} would be created and their corresponding TBB-based \emph{Worker}s are also automatically created for multiple threads' execution.}\par

\begin{figure}[h]
\includegraphics[width=0.98\linewidth, height=13pc]{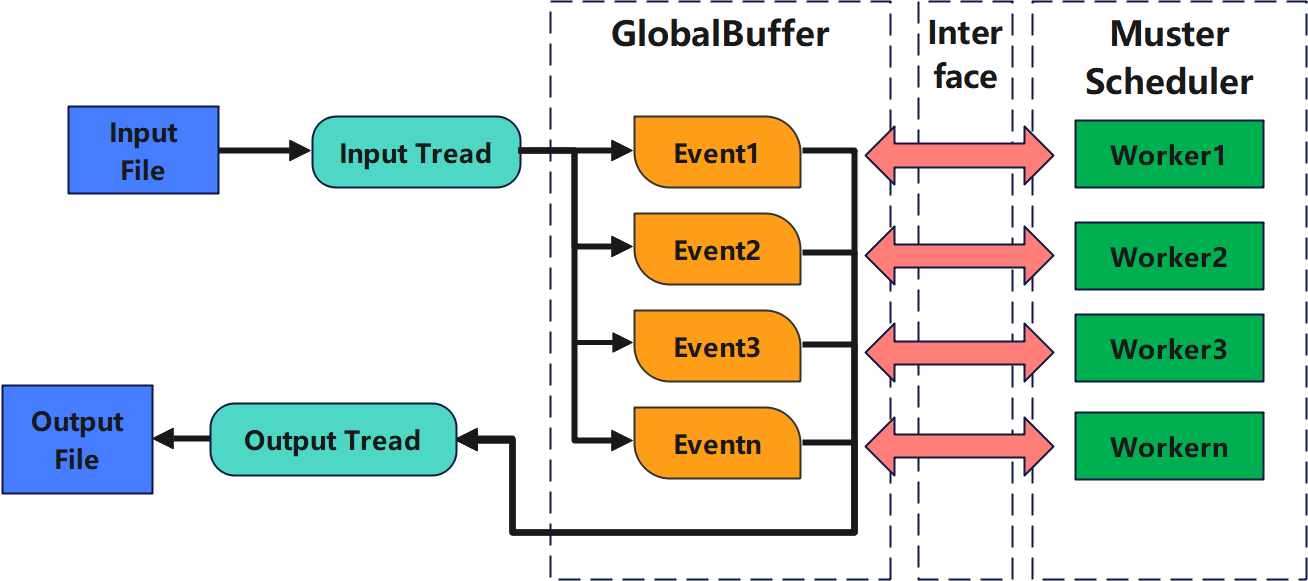}
\caption{\label{label}Event data management for parallel computing.}
\label{fig:parr}
\end{figure}

{Since \emph{Muster} already provides the task scheduler for parallel computing, design and development of an event data management system for parallel computing in podio compatible OSCAR is an important assignment. As illustrated in Section \ref{sec:podio} and \ref{sec:seria}, the event data model for STCF is developed based on podio. However, podio handles event data in memory with \emph{EventStore}, which only holds data objects for one single event. To fulfill requirements of parallel computing, a new mechanism called \emph{GlobalBuffer} is developed to hold multiple transient data objects simultaneously. \emph{GlobalBuffer} manages most of the information which are managed by \emph{EventStore} but for multiple events. Furthermore, it decouples the I/O function from the \emph{EventStore} and assigns reading and writing to two dedicated threads, \emph{Input Thread} and \emph{Output Thread}, respectively.  As illustrated in Figure~\ref{fig:parr},  the \emph{Input thread} would read several event objects and transform them to the \emph{GlobalBuffer} until the \emph{GlobalBuffer} reaches the \emph{Watermark}, a configurable variable to define the number of events. All the events stored in \emph{GlobalBuffer} have \emph{READY} or \emph{FINISH} status. When an event is tagged as \emph{REDAY}, \emph{Muster} scheduler would send it to one dedicated \emph{Worker} for processing. When completed, the event would be tagged as \emph{FINISH}, and finally the \emph{Output Thread} would write the event into a file as persistent data. Then the \emph{GlobalBuffer} would clear the space of this event and wait for next event loop.}\par

{ With this design, a typical sequence of parallel job could be described as follows: There is one dedicated single thread for reading/writing and the transient data to/from a \emph{GlobalBuffer} which simultaneously holds multiple events. When processing, different events are dispatched to different \emph{Worker}s. \emph{Muster} is mainly based on Intel TBB's scheduler, and it creates these \emph{Worker}s and manages their lifetime. Then \emph{Worker}s can be executed in threads concurrently. When a \emph{Worker} is invoked, it grabs and locks an event from \emph{GlobalBuffer} until it is completed. During the execution of a \emph{Worker}, the event is handled by a thread local SNiPER \emph{Task}. This feature ensures that a \emph{Worker} looks the same as a serial SNiPER application. When an event is completed, \emph{Worker}s would try to send it to a dedicated writing thread. As a successful application, parallel simulation based on SNiPER TBB has been adopted in JUNO by integrating Geant4 into \emph{Muster}, which achieves a linear speedup for simulation~\cite{Lin:2017yxy}.}\par

\subsubsection{Detector Data Management Based on DD4hep}
\
\newline

 {To facilitate the detailed detector simulation, a detector simulation framework has been developed as a middleware between Geant4 and SNiPER. It consists of the integration of Geant4 and SNiPER, configurable user interfaces, geometry management and modularized user actions~\cite{Lin:2017yxy}. Based on DD4hep, a geometry management system (GMS)\cite{Li:2021sjr} is  also designed to provide a consistent detector description for simulation, reconstruction and visualization. In GMS, the library of elements and materials is shared and each sub-detector is described in a separated XML file, while the full detector is defined and configured in a mother XML file. Each sub-detector has a version number in order to support different design schemes.}\par

\begin{figure}[h]
\centering
\includegraphics[width=0.98\linewidth]{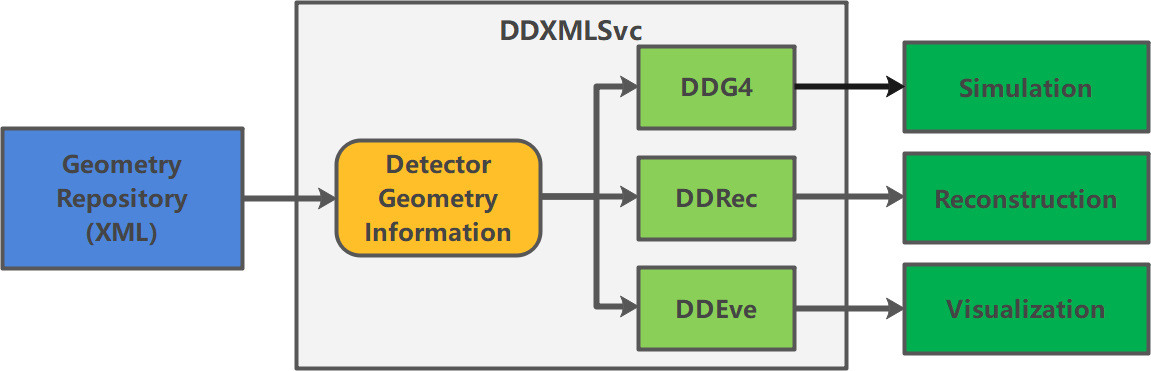}
\caption{\label{label} Data flow of geometry information managed with \emph{DDXMLSvc}. }
\label{fig:ddxmlsvc}
\end{figure}

{A service for detector description in XML format~(\emph{DDXMLSvc}) is developed to manage detector data as a middleware between DD4hep and OSCAR. It manages transformations of detector descriptions for simulation, reconstruction and visualization. When transmitted into \emph{DDXMLSvc}, the detector geometry defined in XML would be parsed and constructed into corresponding sub-detectors or the full detector. Thus, \emph{DDXMLSvc} has all the information of detectors. Information would be delivered to different plugins of DD4hep on demand. For example, information of detector would be delivered to DDG4~\cite{ref4} to transform the geometry into Geant4 format and then used by the detector simulation. Similarly, DDRec~\cite{ref4} for reconstruction and DDEve~\cite{ref4} for visualization are also integrated, as illustrated in Figure~\ref{fig:ddxmlsvc}. With current implementations, OSCAR fully supports simulation, reconstruction and visualization work with one single source geometry definition, which naturally guarantees the consistency of geometry information in different applications.}

\section{Performance\label{sec:perf}}

\indent{Based on the functionality  of the core software, the MC data production chain has been set up, including event generator, full detector simulation and event reconstruction. A typical exclusive process $e^+ e^- \to J/\psi\to\rho(\to\pi^+\pi^-)\pi^0(\to\gamma\gamma)$ is used to commission the underlying algorithms and validate the basic software function. The known decays $J/\psi\to\rho(\to\pi^+\pi^-)\pi^0(\to\gamma\gamma)$ are generated by STCFEvtGen, which is reused from BesEvtGen\cite{Ping:2008zz}. The $J/\psi$ resonance in electron positron collision is simulated by the MC event generator KKMC~\cite{Jadach:1999vf}. Objects of final-state particles, including charged tracks~($\pi^+\pi^-$)and neutral photon showers, are reconstructed with simulated hits through the full reconstruction algorithms.}\par

\begin{figure}[h]
\centering
\subcaptionbox*{(a)}{\includegraphics[width=0.3\textwidth]{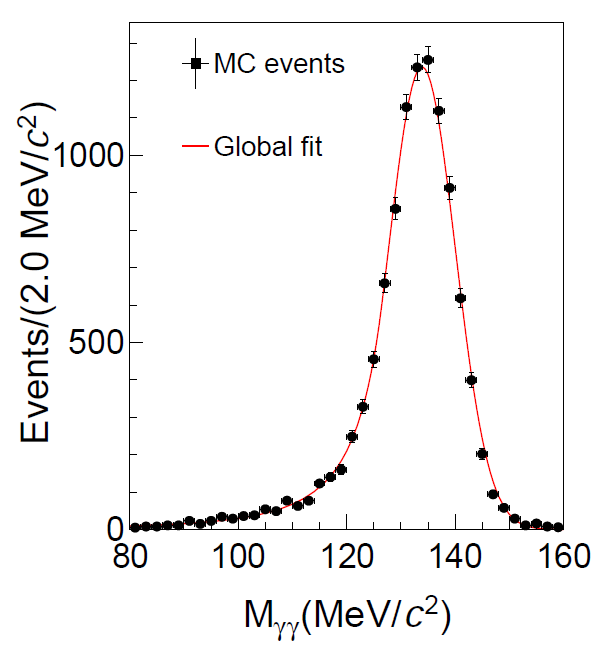}}
\hfill
\subcaptionbox*{(b)}{\includegraphics[width=0.3\textwidth]{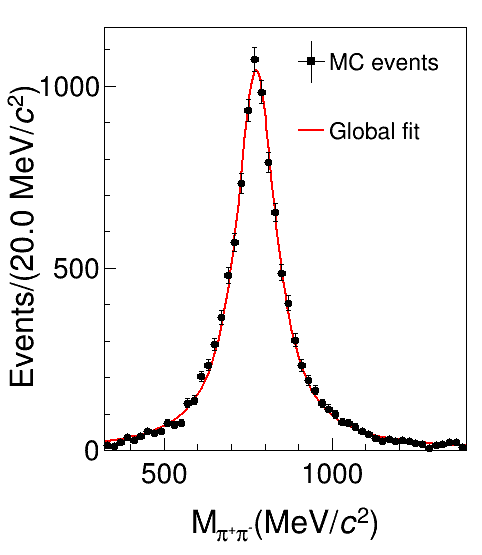}}
\hfill
\subcaptionbox*{(c)}{\includegraphics[width=0.3\textwidth]{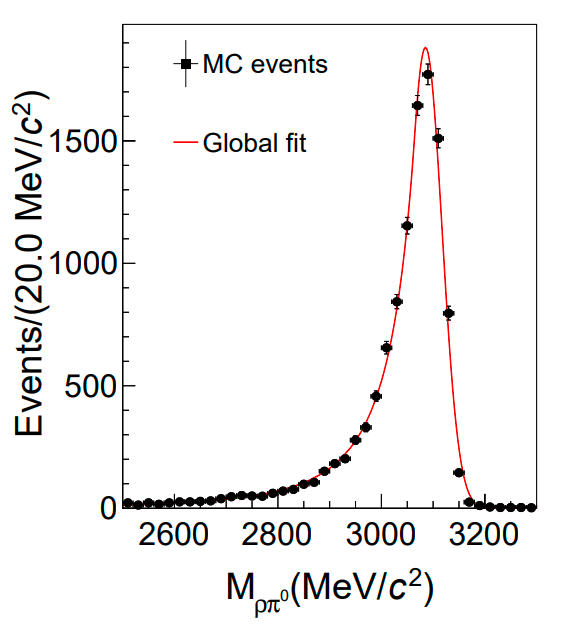}}
\caption{ Plot~(a) shows the invariant mass of reconstructed $\pi^0$ particles~($M_{\gamma\gamma}$), plot~(b) shows the invariant mass of reconstructed $\rho$ particles~($M_{\pi^+\pi^-}$) and plot~(c) shows the invariant mass of reconstructed $J/\psi$ particles~($M_{\rho\pi^0}$) . The dots with error bars indicate MC samples and the red solid lines indicate the fit results of the distribution. In the fit, the $M_{\gamma\gamma}$ distribution in plot~(a) and $M_{\rho\pi^0}$ distribution in plot~(c) are modeled by Crystal-Ball function and the $M_{\pi^+\pi^-}$ distribution in plot~(b) is described with Breit-Wigner function. }
\label{fig:event_rec}
\end{figure}

For each $J/\psi$ candidate, it is required to have two charged tracks with zero net charge and at least two good photons. To suppress final state radiation photons, the angle between photon and the original direction of the nearest charged track must be greater than $15^{\circ}$. The combination with the closest mass to $\pi^{0}$ is considered to be from $\pi^{0}$. The invariant mass distributions of $\pi^{0}$ ($M_{\gamma\gamma}$), $\rho(770)$ ($M_{\pi^+\pi^-}$) and $J/\psi$ ($M_{\rho\pi^0}$) are shown in Figure~\ref{fig:event_rec} (a), (b) and (c), respectively. The Crystal-Ball function is used to describe the $M_{\gamma\gamma}$ and $M_{\rho\pi^0}$ distribution, and the Breit-Wigner function is used to describe the $M_{\pi^+\pi^-}$ distribution. The fitted results are $133.9\pm6.2$ MeV, $3085.2\pm32.1$~MeV and  (M, $\Gamma$) = ($773.6\pm0.9$~MeV, $144.8\pm2.1$~MeV), respectively. In the MC sample, the input values of $\pi^0$, $\rho(770)$ and $J/\psi$ are quoted from PDG~\cite{ParticleDataGroup:2020ssz}, and they are $134.9770\pm0.0005$~MeV, $775.26\pm0.25$~MeV, and $3096.900\pm0.006$~MeV. The reconstructed value and corresponding input value are in good agreement, indicating that the reconstruction of the signal process works well.

At present, a test run of the simulation software is implemented on a dedicated high performance computing cluster under CENTOS 7. Together with the job submission system, large amount of MC samples have been generated, which can be used in further studies of the detector performance and physical potential capabilities.

\section{Conclusion\label{sec:conc}}
  
\indent{In this paper, we present the design and development of the core software of the STCF Offline Data Processing Software, which is based on SNiPER and adopted new technologies and tools, such as DD4hep and podio. Based on current implementations, a chain for full simulation and reconstruction has been set up and large MC samples have been generated for detector performance studies. The study shows the performance of detector simulation and reconstruction could fulfill the requirements of STCF. Meanwhile, OSCAR is also provide a potential solution for other lightweight HEP experiments.}

\acknowledgments
This work was supported by National Natural Science Foundation of China (NSFC) under Contracts Nos. 12025502, 12105158; the international partnership program of the Chinese Academy of Sciences under Grant No. 211134KYSB20200057 ; Double First-Class university project foundation of USTC.\par


\end{document}